\documentclass[reprint,superscriptaddress, amsmath,amssymb, aps, pra, ]{revtex4-2}
\usepackage[verbose]{placeins} 
\usepackage[usenames, dvipsnames]{color}
\usepackage{comment}
\usepackage{hyperref}
\usepackage{subcaption}
\usepackage{float}
\usepackage{ragged2e}
\hypersetup{
    colorlinks=true,
    linkcolor=Blue,
    citecolor=Blue,
    filecolor=Blue,      
    urlcolor=Blue,
    pdftitle={Analytical study of a finite-range impurity in a one-dimensional Bose gas}}
\usepackage{graphicx}
\usepackage{physics}
\usepackage{verbatim}

\def\gIB{g_{\mathrm{IB}}}
\def\aIB{a_{\mathrm{IB}}}

\def\mEff{m_{\mathrm{eff}}}

\begin{document}

\title{Analytical study of a finite-range impurity in a one-dimensional Bose gas}

\author{T. A. Yo\u{g}urt}
 \email{ayogurt@pks.mpg.de}
 \affiliation{%
Max Planck Institute for the Physics of Complex Systems, N\"othnitzer Str. 38, 01187 Dresden, Germany\\
}%

\author{Matthew T. Eiles}
\email{meiles@pks.mpg.de}
 \affiliation{%
Max Planck Institute for the Physics of Complex Systems, N\"othnitzer Str. 38, 01187 Dresden, Germany\\
}%
\date{\today}

\begin{abstract}
One-dimensional Bose gases present an interesting setting to study the physics of Bose polarons, as density fluctuations play an enhanced role due to reduced dimensionality.
Theoretical descriptions of this system have predominantly relied on contact pseudopotentials to model the impurity-bath interaction, leading to unphysical results in the strongly coupled limit. In this work, we analytically solve the Gross-Pitaevskii equation, using a square well potential instead of a zero-range potential, for the ground-state wave function of a static impurity. We compute perturbative corrections arising from infinitesimally slow impurity motion. 
The polaron energy and effective mass remain finite in the strongly coupled regime, in contrast to the divergent behavior obtained using a contact potential. In this limit, we characterize the polaron properties in terms of the dimensionless ratio $\tilde{w}\equiv w/\xi$ between the interaction range $w$ of the impurity-bath potential and the coherence length $\xi$ of the Bose gas. The effective mass exhibits a $1/\tilde{w}$ scaling.  The energy of the attractive polaron scales as $-1/\tilde{w}^3$, whereas the repulsive polaron features subleading corrections to the dark soliton energy at the order $\tilde{w}^3$. 

\end{abstract}

\maketitle

\section{\label{sec: Introduction} INTRODUCTION}

The introduction of atomic impurities into a Bose gas has opened a new avenue for exploring Bose polarons, which were first realized experimentally in a quasi-one-dimensional (1D) system \cite{2012_PRA_Lamporesi_1D_Bose_Polaron_Experiment}. The ability to tune both the sign and the strength of the impurity-bath interaction 
has revealed new regimes that challenge conventional polaron theories \cite{2016_PRL_Bruun_Bose_Polaron_Experiment,2016_PRL_Cornell_Bose_Polaron_Experiment,2025_RPP_Grusdt_Review_Bose_Polaron,2025_Arxiv_Massignan_Review}. Similar to solid-state systems \cite{1954_Frochlich_Electrons_in_lattice_fields,1953_PR_LLP_Transformation,2020_Devreese_Polaron_Lectures}, the leading theoretical framework describes the Bose polaron in terms of phonon excitations of the uniform bath and their interactions with both the impurity and with one another \cite{2014_PRA_Sarma_Chevy_Bose_Polaron,2014_PRA_Demler_Frochlich_Bose_polaron_Coherent_State,2016_Grusdt_New_theoretical_approaches,2016_PRL_Demler_Coherent_State_Beyond_Frohlich,2017_NJP_Grusdt_RG_MF_1D_Bose_Polaron,2018_PRA_Ling_Analytic_1D_Beyond_Frohlich,2018_PRA_Tempere_Bose_Polaron_Ground_State,2022_Atoms_Parish_Repulsive_Polarons_Review}. While this has proved successful in the weak and intermediate coupling regimes, it is inadequate at strong coupling due to the neglect of density modulations around the impurity \cite{2020_PRR_Fleisheur_Exact_1D_Bose_Polaron,2021_PRL_Massignan_GP_Bose_Polaron,2022_PRA_Nikolay_Strongly_Interacting_Impurities,2023_PRA_Nikolay_EffMass_Bose_Polaron,alhyder2025lattice,hartweg2025bose}. These modulations become even more pronounced in lower dimensions \cite{2003_Book_Giamarchi_1D_QP,2011_RMP_1D_Bose_Gas_Review,2025_RPP_Grusdt_Review_Bose_Polaron}, as evidenced by the fact of vanishing quasiparticle weight \cite{2017_NJP_Grusdt_RG_MF_1D_Bose_Polaron} and impurity self-trapping for any repulsive interaction in 1D at the thermodynamic limit \cite{2006_PRL_Timmermans_Self_Trapping,2008_EL_Self_trapping_impurity,2006_PRA_Timmermans_Self_Localized_Impurity_1d,2025_PRA_Fleuschauer_Impurity_1D_Selflocalization}.

To account for these density modulations near the impurity, the exact mean-field (MF) solutions to the Gross-Pitaevskii (GP) equation have been obtained \cite{2017_PRA_Volosniev_Analytical_1D_Polaron,2020_PRR_Fleisheur_Exact_1D_Bose_Polaron} by treating the impurity-bath interaction as a 
contact pseudopotential with the coupling strength $\gIB \propto -1/\aIB$, where $\aIB$ is the zero-energy scattering length. This framework has been applied to study both the properties of a single Bose polaron \cite{2020_PRR_Fleisheur_Exact_1D_Bose_Polaron} and the interactions between two impurities \cite{2021_PRL_Michael_Polaron_1D,2022_PRA_Petkovic_Polaron_Interactions}. Remarkably, the analytical results have shown strong agreement -- even at the MF level-- with Monte Carlo simulations. 
However, in the strongly coupled limit $\gIB \to \pm \infty$, the MF results yielded diverging values of key polaron properties, including the energy of the attractive impurity and the effective mass for both attractive and repulsive cases. The Monte Carlo results displayed a similar unbounded growth as the number of particles included in the simulations increased \cite{2017_PRA_Parisi_QMC_1D_Polaron_,2021_PRL_Michael_Polaron_1D}. This divergent behavior can be attributed to the fact that the pseudopotential itself diverges in the strong-coupling limit, inadequately representing the actual short-range impurity–bath interaction. 

In this Article, we investigate how a finite-range potential, namely, a square well, corrects this behavior in the strong coupling limit. Within the MF framework, we obtain semi-analytical ground-state solutions to the GP equation.
Although this approach has been successfully employed to obtain physical results at strong coupling for the Bose polaron in 3D \cite{2022_SciPost_Schmidt_GP_Bose_Polaron,2021_PRL_Massignan_GP_Bose_Polaron}, it has not yet been employed in 1D \cite{casteels2012polaronic,2015_PRA_Impurity_in_1D_Bose_Gas,2016_PRA_Pelster_Localized_impurity_in_BEC,2017_PRA_Volosniev_Analytical_1D_Polaron,pastukhov2017impurity,2017_NJP_Grusdt_RG_MF_1D_Bose_Polaron,2020_PRR_Mistakidis_1D_Bose_Gas_Impurity,2021_PRA_Ristivojevic_Exact_1D_Bose_Gas,2019_AP_Pastukhov_MF_Bose_Polaron_1D_spectrum,abdullaev2020bosonic,leboeuf2001bose,2025_Arxiv_Petkovic_Relaxation_Dynamics,2025_PRA_Fleuschauer_Impurity_1D_Selflocalization}. Additionally, we compute the first-order corrections to the wavefunctions arising from the impurity’s infinitesimally slow motion. We compute the polaron energy and effective mass as a function of the coupling strength $\gIB$. In the weak to intermediate coupling regime, we observe that the polaron properties exhibit minimal sensitivity to the short-range character of the impurity interaction and closely align with those predicted for a contact potential. In contrast, in the strong coupling limit, $\gIB \to \infty$, the polaron energy and effective mass remain finite and become analytical functions only of the dimensionless ratio of the interaction range relative to the Bose gas coherence length, $\tilde{w} \equiv w/\xi$.
 For attractive polarons, the energy scales as $-1/\tilde{w}^3$, while for repulsive polarons, we compute subleading corrections of order $\tilde{w}^3$ to the dark soliton energy. The effective mass in both cases scales as $1/\tilde{w}$.

This paper is organized as follows. Section \ref{sec: Theory} explains the general theoretical framework and the solutions to the GP equation. The polaron properties of the finite-ranged impurity are discussed in Section \ref{sec: Results}. Section \ref{sec: Universality} presents the characterization of the strong coupling limit. Finally, we conclude in Section \ref{sec: Conclusion}.

\section{\label{sec: Theory} THEORY}

We consider a mobile impurity of mass $m_I$ immersed in a 1D gas of bosonic atoms of mass $m$. The intraspecies interaction is described by a contact pseudopotential with coupling strength $g$. The Hamiltonian of such an impurity-bath system with the chemical potential $\mu$ is represented in the second-quantized formalism as
\begin{eqnarray} \label{Hamiltonian_Lab_Frame}   
    \hat{\mathcal{K}} = \frac{\hat{p}_I^2}{2m_I}+ \int dx \hat{\Psi}^{\dagger}(x)&\bigg[& -\frac{\hbar^2 \hat{\partial}_x^2}{2m} + \frac{g}{2} \hat{\Psi}^{\dagger}(x)\hat{\Psi}(x)\nonumber  \\ &+&{U}_{\mathrm{IB}}(x-\hat{x}_I)-\mu\bigg]\hat{\Psi}(x),\nonumber \\
\end{eqnarray}
where $\hat{\Psi}(x)$ is the field operator of the bath particle at position $x$. ${U}_{\mathrm{IB}}(x)$ denotes the two-body potential between bath particles and the impurity localized at the position $\hat{x}_I$. The operator $\hat{p}_I$ denotes the momentum of the impurity.

In the absence of an impurity, Lieb and Liniger \cite{1963_Lieb_Exact} demonstrated that the system~\eqref{Hamiltonian_Lab_Frame} admits an exact solution by Bethe ansatz, characterized by the gas parameter, $\gamma \equiv mg/n$, where $n$ is the gas density. Despite the absence of long-range coherence in an infinite 1D Bose gas, the GP ansatz accurately describes the ground state and excitations when $\gamma \lesssim 2$ \cite{1963_Lieb_Exact_2}. Moreover, studies of the 1D polaron problem \cite{2020_PRR_Fleisheur_Exact_1D_Bose_Polaron,2021_PRL_Michael_Polaron_1D, 2025_RPP_Grusdt_Review_Bose_Polaron} found strong agreement between the GP theory and Monte Carlo simulations for sufficiently weak intraspecies interactions. In this work, we likewise focus on weakly interacting bosons and develop the GP framework for a finite-range impurity.

\subsection{Gross-Pitaevskii Equation}
 We move to the frame of the impurity by applying the Lee-Low-Pines coordinate transformation by $\hat{\mathcal{K}}_{\mathrm{LLP}} = \hat{\mathcal{S}}^{-1} \hat{\mathcal{K}} \hat{\mathcal{S}}$, where $\hat{\mathcal{S}} = \exp{-i \hat{x}_I  \hat{p}_B/\hbar}$, and  $\hat{p}_B = -i \hbar\int dx  \hat{\Psi}^{\dagger}(x)  \partial_x \hat{\Psi}(x)$ is the total phonon momentum. This transformation results in a Hamiltonian:
\begin{widetext}
\begin{eqnarray} \label{Hamiltonian_LLP}   
    \hat{\mathcal{K}}_{\mathrm{LLP}} = \frac{(p_{\mathrm{tot}}-\hat{p}_B)^2}{2m_I} + \int dx \hat{\Psi}^{\dagger}(x) \bigg[ -\frac{\hbar^2 \hat{\partial}_x^2}{2m_r} + \frac{g}{2} \hat{\Psi}^{\dagger}(x)\hat{\Psi}(x)  +{U}_{\mathrm{IB}}(x)-\mu\bigg]\hat{\Psi}(x),
\end{eqnarray}    
\end{widetext}
where $m_r \equiv m_I m/(m_I +m)$ is the reduced mass. The total momentum $p_{\mathrm{tot}}\mathbf{e_x}$ of the impurity–bath system is conserved.
We proceed by applying the MF approximation to the Hamiltonian in  \eqref{Hamiltonian_LLP}, replacing the operators by their expectation values, $\hat{\Psi}(x) \simeq \langle \hat{\Psi}(x) \rangle =  \Psi(x)$, while neglecting any fluctuations. We apply the variational analysis $\frac{\delta \mathcal{K}_{\mathrm{LLP}}}{\delta \Psi^*}= 0$ to derive the time-independent GP equation governing the Bose field with non-zero total system momentum $p_{\mathrm{tot}}$:
\begin{eqnarray}\label{GPE_Finite_Momentum}
    \mu \Psi(x) &=& -\frac{\hbar^2 \partial_x^2 \Psi(x)}{2m_r} + U_{\mathrm{IB}}(x) \Psi(x) +g|\Psi(x)|^2 \Psi(x) \nonumber\\ &+& \frac{i \hbar \partial_x \Psi(x)\big[p_{\mathrm{tot}} -\int dx'\Psi^*(x') (-i\hbar) \partial_{x'} \Psi(x')\big]}{m_I}, \nonumber \\
\end{eqnarray}
where the term in square brackets in the second line is the impurity momentum $p_I= p_{\mathrm{tot}}-p_B$ at the MF level, with ${p}_B = \langle\hat{p}_B\rangle$. 

We set the chemical potential $\mu = gn_0$, where $n_0$ is the uniform bath density far from the impurity. 
The length scale $\tilde{\xi} = \xi \sqrt{\frac{m}{m_r}}$ is set by the coherence length $\xi = \frac{\hbar}{\sqrt{2m\mu}}$. The dimensionless GP equation governing the Bose field $\phi(x) \equiv \Psi(x)/ \sqrt{n_0}$ is
\begin{equation}\label{GPE_Dimensionless}
    -\partial_{{x}{x}} \phi({x}) + \big[|\phi({x})|^2+ \tilde{U}_{\mathrm{IB}}({x})-1\big] \phi({x}) = -i \tilde{v} \partial_{x} \phi(x). 
\end{equation}
Here, $\tilde{v} = 2 {m_r\tilde{\xi}}v/{\hbar}$ is the dimensionless impurity speed with $v = {p_{{I}}}/{m_I}$. 
The dimensionless potential is $\tilde{U}_{IB} (x)= \frac{{U}_{IB}(x)}{gn_0}$. 
\subsection{Two-body Potential}
We model $U_{\mathrm{IB}}(x)$ with a square well
\begin{equation}
    U_{\mathrm{IB}}(x) = V_0 {\Theta}(w/2-|x|),
\end{equation}
because it admits an analytical solution and has a finite range.
${\Theta}(x)$ denotes the Heaviside step function, and $V_0$ and $w$ are the well's depth and width, respectively. 
The zero-energy scattering length, $\aIB$, corresponding to the even-parity solution of the two-body problem is known analytically:
\begin{equation}
    \frac{\aIB}{w/2} =  1+ \frac{\cot{z}}{z},
    \label{aIB_w_V0_Att}
\end{equation}
\begin{equation}
    \frac{\aIB}{w/2} =  1- \frac{\coth{z}}{z},
    \label{aIB_w_V0_Rep}
\end{equation}
for $V_0<0$ and $V_0>0$ \cite{2000_EJP_1D_Scattering,2010_CJP_FiniteWell_Scattering_Length}, respectively. Here $z = \sqrt{\frac{|{V}_0| 2m_r}{\hbar^2}}\frac{w}{2}$. In 1D, the relation between the coupling strength $\gIB$ and the scattering length $\aIB$ is 
given by $\gIB = -\frac{\hbar^2}{m_r \aIB}$. 
(See Appendix \ref{App:ScatteringLength} for details.)

\subsection{Calculation of Observables}
Two key observables are the polaron energy and the effective mass of the impurity. While the polaron energy can be obtained by solving \eqref{GPE_Dimensionless} for the ground-state wave function with zero total momentum, $\phi_0(x)$, evaluating the effective mass requires solutions at $p_{\mathrm{tot}} \neq 0$. To this end, we perform a low impurity velocity expansion of the solution $\phi(x)$ of the GP equation \eqref{GPE_Dimensionless} around $\phi_0(x)$ in the form
\begin{equation} \label{phi0_Corrections}
    \phi(x) \simeq \phi_0(x) + i\tilde{v} \phi_1(x),
\end{equation}
where $|\tilde{v}|\ll 1$. 
Upon inserting Eq.~\eqref{phi0_Corrections} into Eq.~\eqref{GPE_Dimensionless}, we obtain 
\begin{eqnarray}\label{GP_Correction_Equations}
    0 &=& -\partial_{xx}\phi_0 + \big(\phi_0^2-1 \big)\phi_0  + \tilde{U}_{IB}(x) \phi_0,\label{phi0_equations} \\
    -\partial_x \phi_0 &=& -\partial_{xx}\phi_1 +\big(\phi_0^2-1 \big)\phi_1  + \tilde{U}_{IB}(x) \phi_1,
    \label{phi1_equations}
\end{eqnarray}
 to first order in the impurity velocity $\tilde{v}$ \cite{2023_PRA_Nikolay_EffMass_Bose_Polaron}.
The non-uniform phase $\theta(x)$ of the moving condensate is determined by $\phi_1$, via $\theta(x) \equiv \arctan{[\tilde{v} \phi_1(x)/\phi_0(x)]}$. 
Once the expansion \eqref{phi0_Corrections} is replaced in the total energy functional \eqref{Hamiltonian_LLP},  both the ground-state polaron energy and the correction due to a small impurity velocity, i.e., $E = E_{\mathrm{pol}} + \frac{1}{2}m_{\mathrm{eff}} v^2$, can be computed, yielding
\begin{eqnarray}
    \frac{E_{\mathrm{pol}}}{gn_0^2\tilde{\xi}} 
    &=& \frac{1}{2} \int dx (1-|\phi_0|^4), \label{pol_energy}\\
    \frac{m_{\mathrm{eff}}}{m_I} &=& 1 - \frac{4n_0 \tilde{\xi}m_r}{m_I}  \int dx \phi_1 \partial_x \phi_0,\label{pol_eff_mass} 
\end{eqnarray}
where $\phi_1$ is fixed to be a continuous function of $x \in (-\infty, \infty)$ and $\mEff$ is the effective mass of the impurity.  Note that $E_{\mathrm{pol}}$ is the difference between the ground-state energies of the many-body system with and without the impurity \footnote{Note that one also needs to subtract the interaction energy between the background particles of the uniform density $\sqrt{n_0}$ and the particles repelled (attracted) from (into) the vicinity of the impurity.}. 
\begin{figure}[t]
    \begin{subfigure}{0.238\textwidth}
        \includegraphics[width = \textwidth,keepaspectratio]{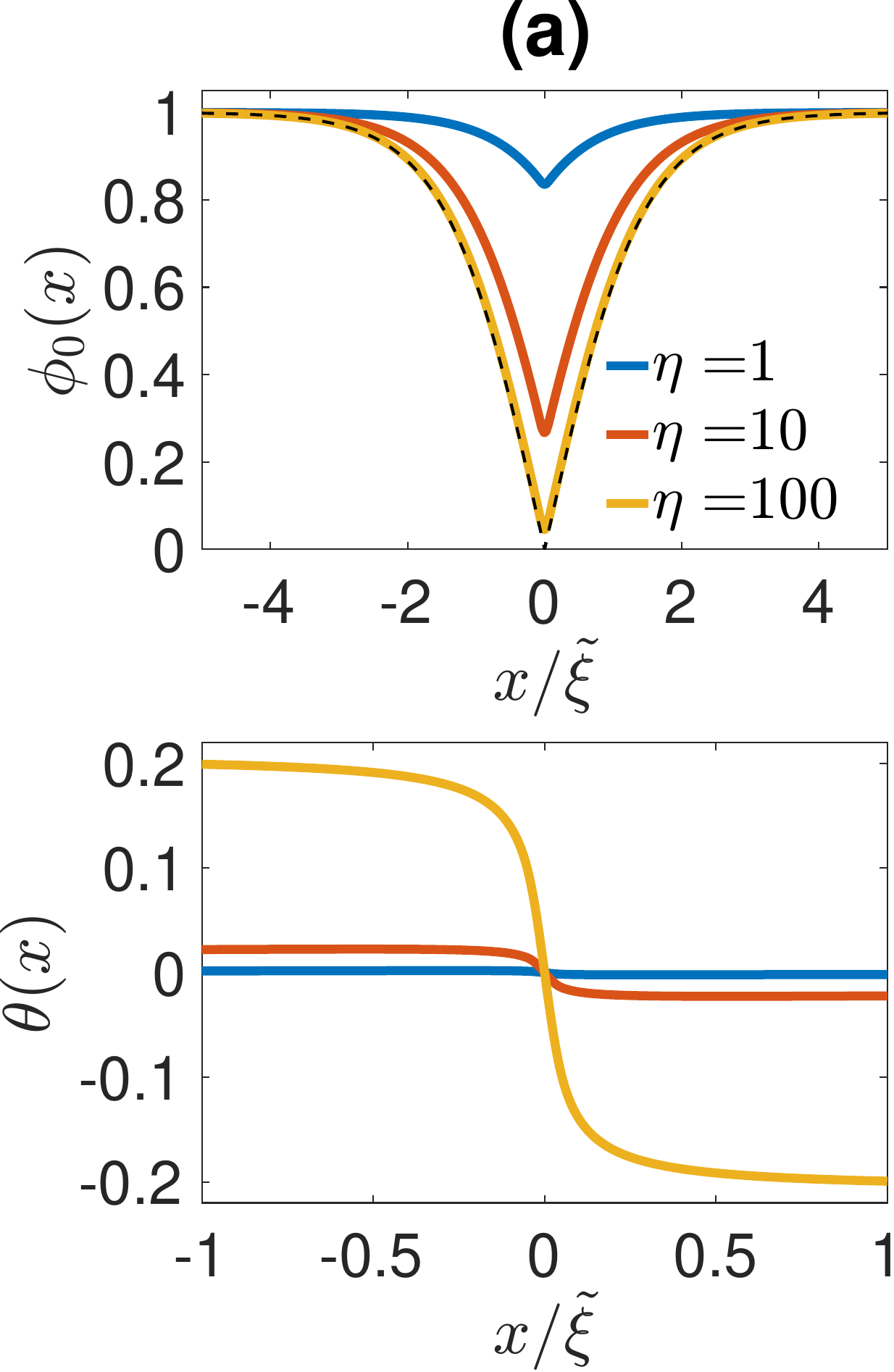}
    \end{subfigure}
    \begin{subfigure}{0.23\textwidth}
        \includegraphics[width = \textwidth,keepaspectratio]{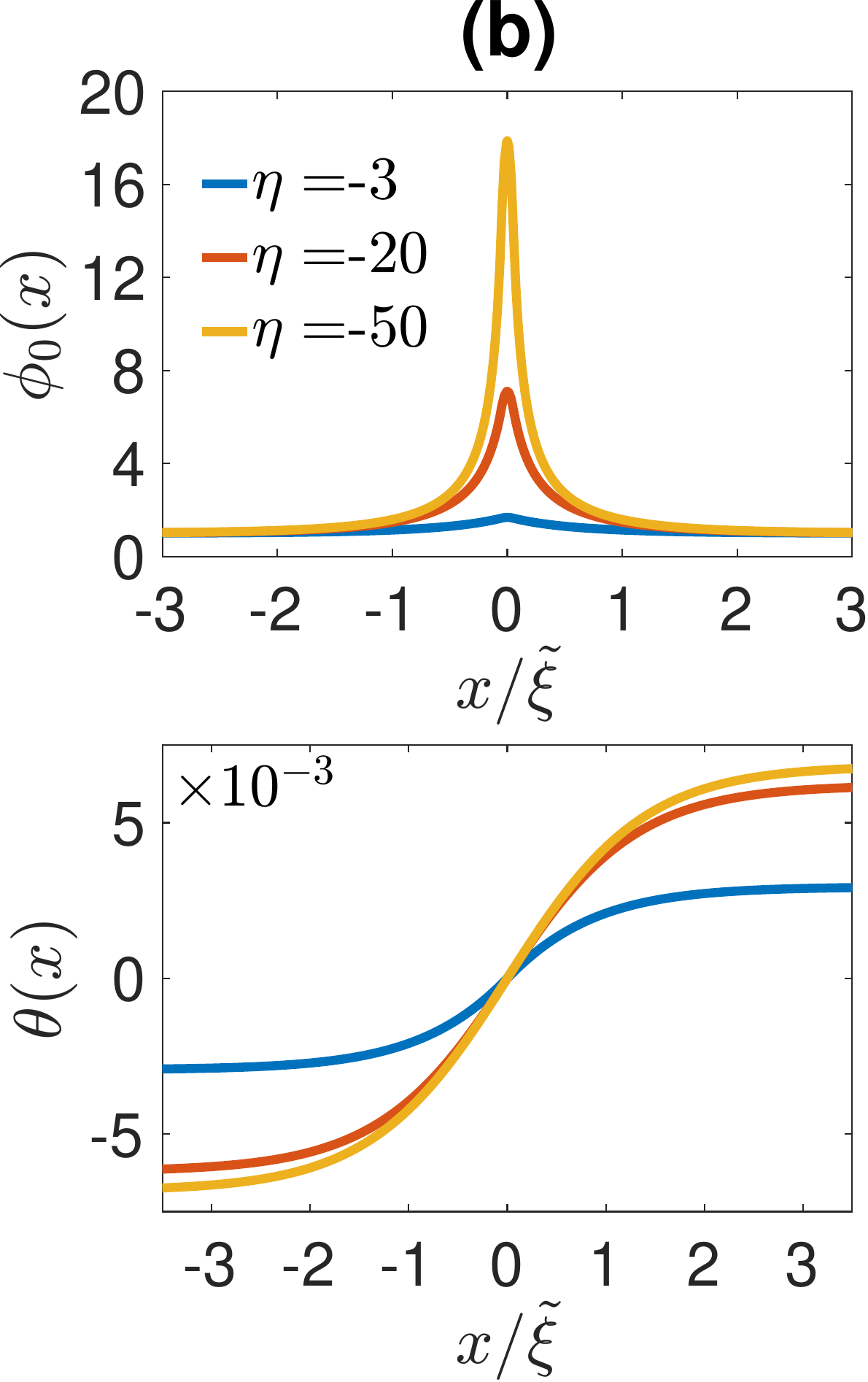}
    \end{subfigure}
    \caption{\justifying The ground-state wave function $\phi_0({x})$ (upper panels) and the corresponding phase function $\theta(x) \equiv \arctan[\tilde{v}\phi_1({x})/\phi_0({x})]$ (lower panels)  of the \textbf{(a)} repulsive and \textbf{(b)} attractive interactions for various coupling strengths $\eta\equiv \gIB/g$. The well size is fixed as $w = 0.1\tilde{\xi}$. The velocity is fixed as $\tilde{v} = 0.01$. The thin dashed line in the upper panel of (a) represents the soliton-like profile $\tanh(|x|/(\sqrt{2}\tilde{\xi}))$, shown for comparison. 
    }
    \label{fig:wavefunctions}
\end{figure}
\subsection{Semi-analytical Solution of Ground-State Wavefunction}
We restrict our consideration to the ground state solution, which must satisfy $\phi_0(-x)=\phi_0(x)$,  $\partial_x\phi_{0}(0)=0$,  $\phi_{0}(\pm \infty)=1$, and (along with its derivative) must be continuous at the edges of the square well. 
As the specific solution depends on the sign of $V_0$, we treat each case separately, beginning with $V_0<0$. The semi-analytical solution for $\phi_0(x)$ can be obtained by integrating Eq.~\eqref{GPE_Dimensionless} twice. The first integration yields the constant of motion $K = (\partial_x \phi_0)^2/2 + V[\phi_0]$, where $V[\phi_0] = -\phi_0^4/4 + W_a \phi_0^2/2$ and $W_a = |\tilde{V}_0|+1$ with $\tilde{V}_0 = V_0/(gn_0)$. We perform the second integration by putting the GP equation into the form $\partial_x \phi_0 =  \sqrt{2(K-V[\phi_0])}$, which can be related to Jacobi elliptic functions \cite{1948_Abramowitz_Mathematics,2013_Lawden_Elliptic_Functions} (See Appendix. \ref{Appendix:Jacobi}). The solutions inside and outside the square well are
\begin{equation}
    \phi_{0}\left(x\right) = 
    \begin{cases}
    \tanh^{-1}{\left(\frac{|x|+x_a}{\tilde{\xi}\sqrt{2}}\right)},& |x|>\frac{w}{2}, \\
    \sqrt{\frac{2W_a\nu_a}{\nu_a + 1}} \mathrm{cd}\left( \sqrt{\frac{W_a}{\nu_a + 1}} \frac{x}{\tilde{\xi}},\nu_a\right), &  |x|<\frac{w}{2}, 
    \end{cases}
    \label{wf_att_inside}
\end{equation}
where $\mathrm{cd}(x,\nu_a)$ is a Jacobi elliptic function and $\nu_a$ is the square of the elliptic modulus. The solution exists for $\nu_a \in [0,1]$. 
Imposing the boundary conditions leads to a transcendental equation (see \eqref{BC_attractive} in Appendix \ref{Appendix:BC})  determining the value of $\nu_a$. When multiple roots exist, the root with the largest value of $\nu_a$ corresponds to the lowest energy solution \footnote{The function $\mathrm{cd}(x,\nu)$ interpolates between the $\cos(x)$ in the limit $\nu \to 0$ and a constant, in the limit $\nu \to 1$. The ground state solution is given by the larger elliptic parameter $\nu$, corresponding to a nodeless wavefunction. The solution with the smaller $\nu$ corresponds to an excited state solution, with at least one node in the range of the impurity. }. 
\begin{figure}[t]
    \begin{subfigure}{0.239\textwidth}
        \includegraphics[width = \textwidth]{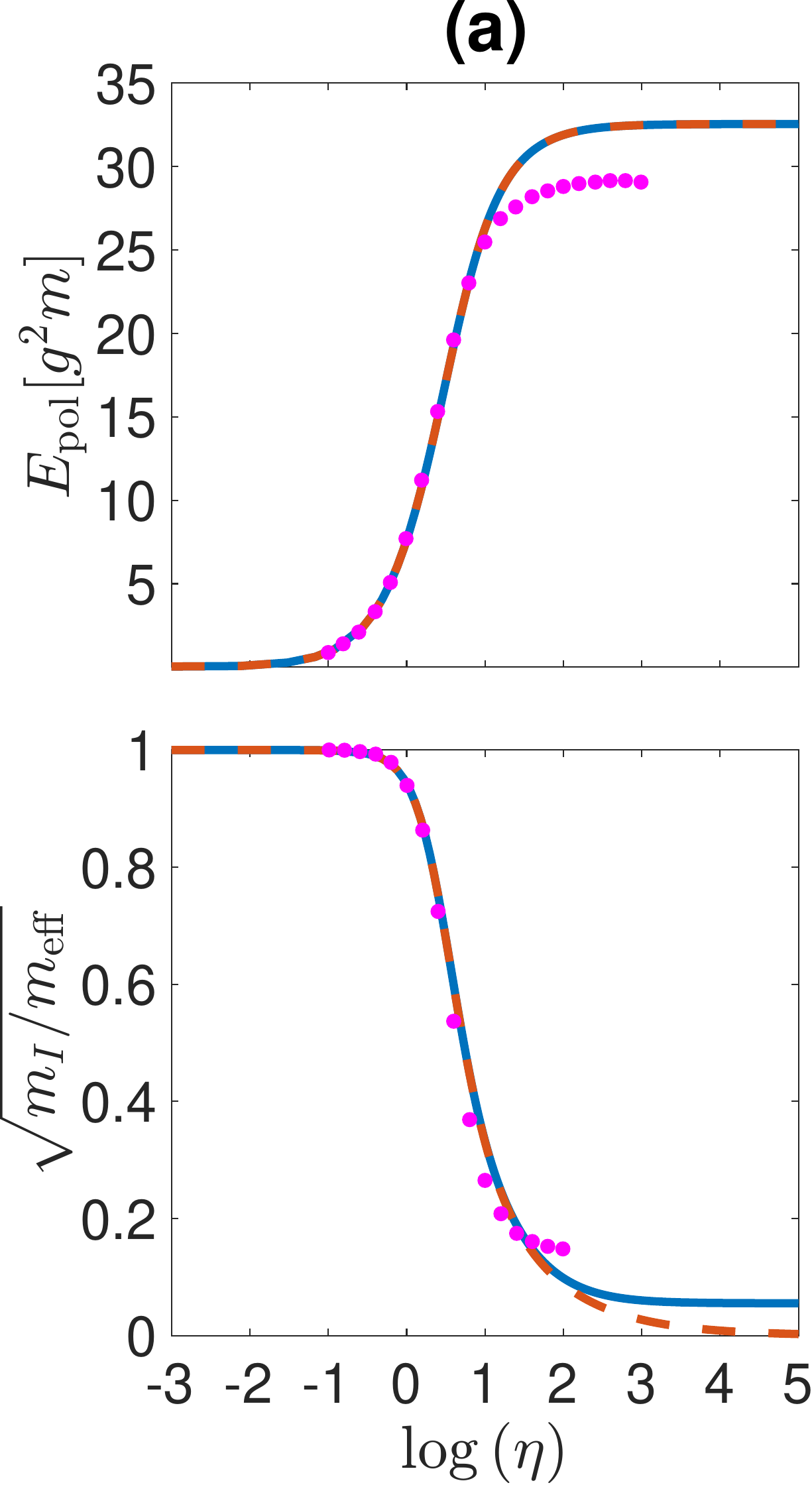}
    \end{subfigure}
    \begin{subfigure}{0.238\textwidth}
        \includegraphics[width = \textwidth]{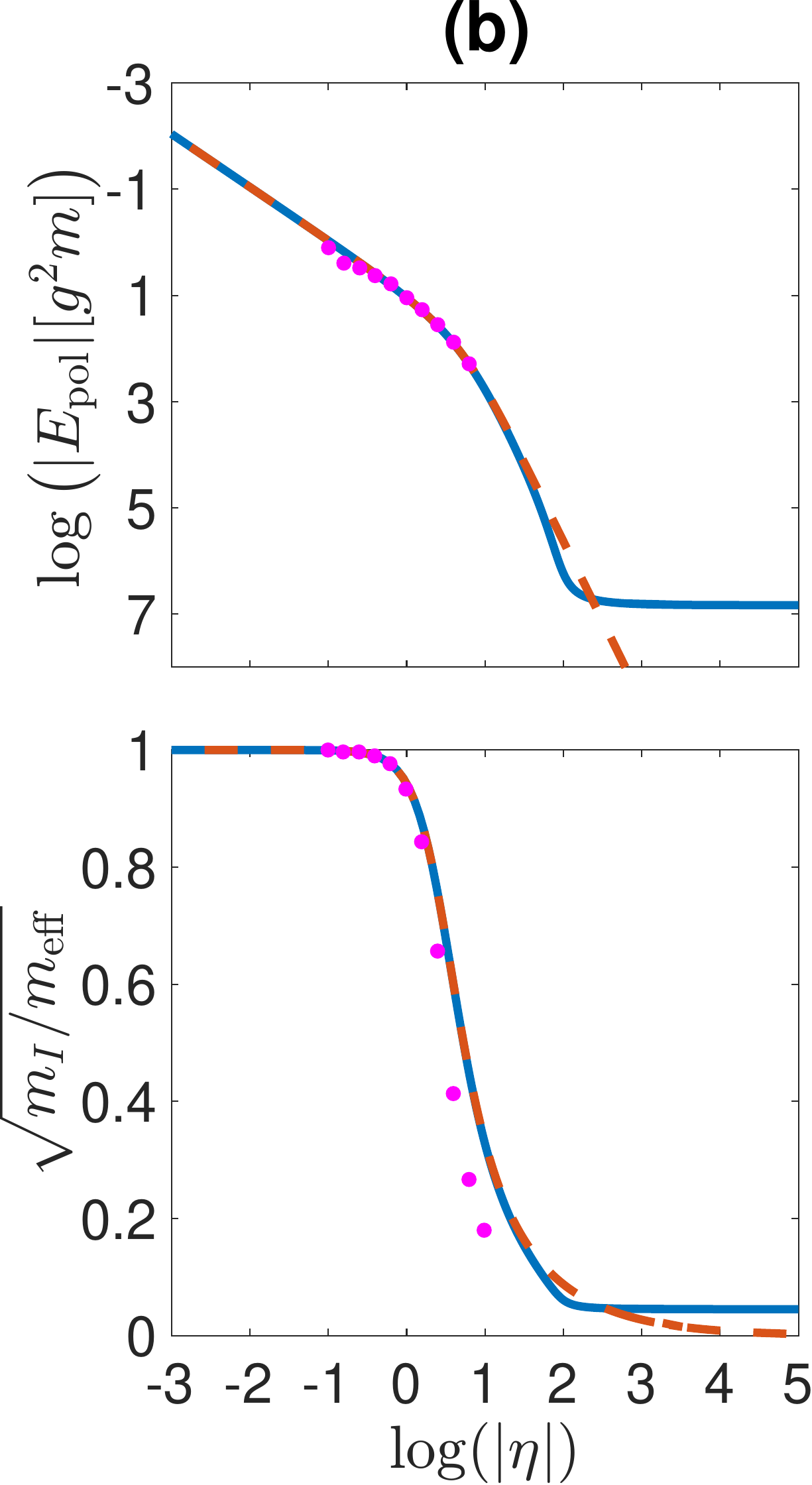}
    \end{subfigure}
    \caption{\justifying The polaron energy and effective mass of an impurity assuming a \textbf{(a)} repulsive and \textbf{(b)} attractive square-well potential (blue solid line), and contact potential (dashed red line) as a function of the coupling strength $\eta \equiv g_{\mathrm{IB}}/g_{\mathrm{BB}}$. The width is again $w = 0.1\xi$. The magenta points show the DMC results obtained for a system of 100 particles with a contact impurity-bath potential \cite{2020_PRR_Fleisheur_Exact_1D_Bose_Polaron}.
    }
    \label{fig:Polaron_Properties}

\end{figure}

When $V_0>0$, the solution for $|x|>w/2$ is
\begin{eqnarray}
    \phi_{0}\left(|x|>\frac{w}{2}\right) = \tanh{\left(\frac{|x|+x_r}{\tilde{\xi}\sqrt{2}}\right)}.
    \label{wf_rep_outside}
\end{eqnarray}
The case $x_r = 0$ corresponds to the dark soliton profile, arising in the strongly repulsive zero-range limit where the vanishing density at the impurity position enforces a solitonic surrounding. 

Within the well ($|x|<w/2$), the solution depends on the sign of $W_r \equiv \tilde{V}_0-1$:
\begin{equation}
\phi_{0}\left(x\right) = 
\begin{cases}
\sqrt{\frac{2W_r (1-\nu_r)}{2\nu_r -1}} \mathrm{nc}\left( \sqrt{\frac{W_r}{2\nu_r-1}}\frac{x}{\tilde{\xi}},\nu_r\right) & W_r > 0, \\
\sqrt{\frac{2|W_r| }{\nu_r +1}} \mathrm{dc}\left( \sqrt{\frac{|W_r|}{\nu_r+1}}\frac{x}{\tilde{\xi}},\nu_r\right) &  W_r < 0. 
\end{cases}
\label{wf_rep_inside}
\end{equation}
Here, $\mathrm{nc}(x,\nu_r)$ and $\mathrm{dc}(x,\nu_r)$ are Jacobi elliptic functions, and $\nu_r$ is determined numerically after imposing boundary conditions (see Eq.~\eqref{BCs_Repulsive1} and \eqref{BCs_Repulsive2}).

Some clarification regarding the boundary conditions for $\phi_1(x)$ is necessary, as the phase $\theta(x)$ of the GP wavefunction is only defined up to an arbitrary global phase. First, since the spatial derivative of the phase function corresponds to the velocity field of the Bose gas, $\phi_1(x)$ is assumed to converge to a constant at infinity, i.e. $\partial_x \phi_1(x\to \pm \infty) = 0$ to ensure that there is no flow far away from the impurity. Second, due to the gauge freedom in the definition of the phase, $\phi_1(x)$ can be shifted by an arbitrary multiple of $\phi_0(x)$. Without loss of generality, one can fix this freedom by imposing the condition $\phi_1(0) = 0$. Together, these two conditions uniquely determine $\phi_1(x)$ for a given $\phi_0(x)$.

\section{\label{sec: Results}RESULTS and DISCUSSION}

We now present the polaron energy and effective mass for a short-ranged impurity, $w/\tilde{\xi} \ll 1$.  Relevant system parameters are constrained to those used in the one-dimensional polaron experiment \cite{2012_PRA_Lamporesi_1D_Bose_Polaron_Experiment}, with the gas parameter fixed at $\gamma = 1/(2n_0^2\xi^2)=0.438$, and the impurity-bath mass ratio set to $m_I/m = 0.47$.

In Fig.~\ref{fig:wavefunctions}, we show the ground-state wave function $\phi_0(x)$ and the phase function $\theta(x)$ for both repulsive and attractive interactions. The width of the square well is fixed to $w = 0.1 \tilde{\xi}$, while its depth $V_0$ is varied to give different coupling strengths $\eta \equiv \gIB/g$. 
As the magnitude of the coupling strength $|\eta|$ increases, the density at the impurity decreases (increases) for the repulsive (attractive) impurity. 
For the repulsive impurity at strong coupling strength, the density profile increasingly resembles, but never fully converges to, that of a dark soliton.

Two features should be highlighted in the behavior of $\theta(x)$. First, its derivative $\partial_x \theta(x)$, corresponding to the fluid flow field of the Bose gas, acquires a negative (positive) sign in the presence of a repulsive (attractive) impurity. The sign inversion in the case of the repulsive impurity can be understood as a consequence of the Bose gas flowing in the $-x$ direction in response to the impurity moving in the $+x$ direction. Second, for the repulsive impurity, the phase contrast between $x \to \pm \infty$ is roughly associated with the ratio of the dip density $n_{min}$ and the uniform density $n_0$, similar to the solitons satisfying $\theta(+\infty)-\theta(-\infty) \propto -\cos^{-1} \left( \sqrt{\frac{n_{min}}{n_0}}\right)$ \cite{pethick2008bose}. 

In Fig.~\ref{fig:Polaron_Properties}, we show the polaron energy and the effective mass over a broad range of coupling strengths $|\eta| \in [10^{-3},10^5]$. Our results for a square well potential (blue solid curves) are compared with those obtained for a contact potential using either a MF treatment of the GP theory (red dashed lines) \cite{2020_PRR_Fleisheur_Exact_1D_Bose_Polaron} or diffusion quantum Monte Carlo (DMC)  (scattered points) \cite{2020_PRR_Fleisheur_Exact_1D_Bose_Polaron,2021_PRL_Michael_Polaron_1D}. 
The energy of the repulsive square well impurity coincides precisely with that of the contact potential for all $\eta$, and asymptotically approaches the energy of a dark soliton in the strong-coupling regime. The effective mass of the square-well impurity also agrees with the contact impurity for couplings $\eta \lesssim 100$; however, it saturates to a finite value at the strong-coupling limit $\eta \to \infty$ only for the finite-size impurity. This saturation is even more apparent for the attractive impurity. In this case, both the polaron energy and the effective mass initially agree with the contact potential results, but begin to deviate around $|\eta| \simeq 50$. While the polaron energy for the contact polaron diverges as $\gIB \to \infty$, the energy in the case of a finite-range potential saturates at a finite value. 

The convergence to a finite value in the strong coupling limit can be understood by considering the relation between the finite-range potential and the corresponding coupling strength $\gIB \propto -\aIB^{-1}$ of the pseudopotential. A divergent $\gIB$ corresponds to a finite-valued finite-range potential, whose zero-energy scattering length satisfies $\aIB \to 0$ (See Fig.~\ref{fig:aIB_gIB}). Therefore, in this limit, the pseudopotential provides a poor representation of the actual impurity potential. 
 
 \begin{figure}
    \begin{subfigure}{0.235\textwidth}
        \includegraphics[width = \textwidth,height=1.43\textwidth]{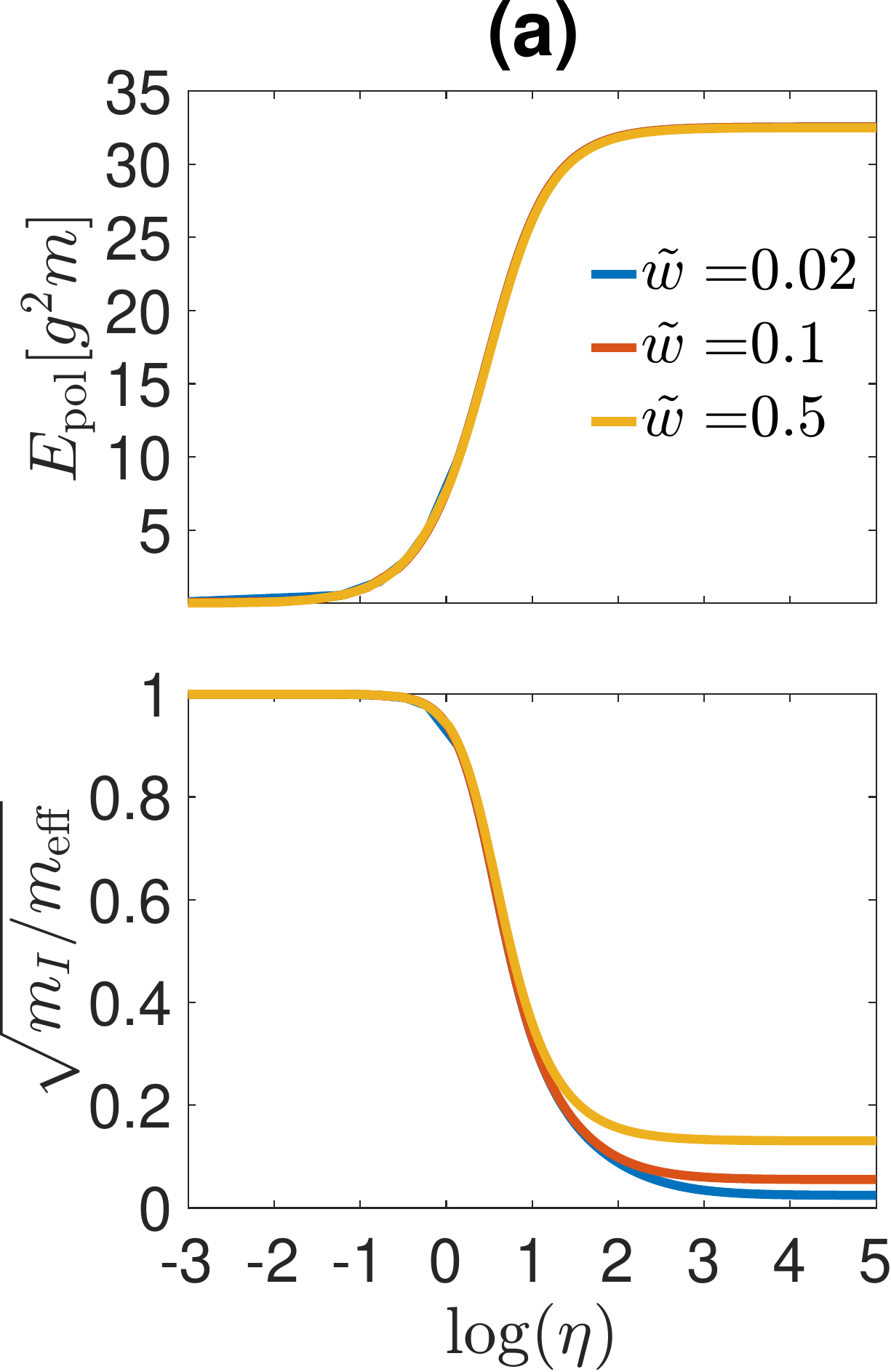}
    \end{subfigure}
    \begin{subfigure}{0.235\textwidth}
        \includegraphics[width = \textwidth,height=1.43\textwidth]{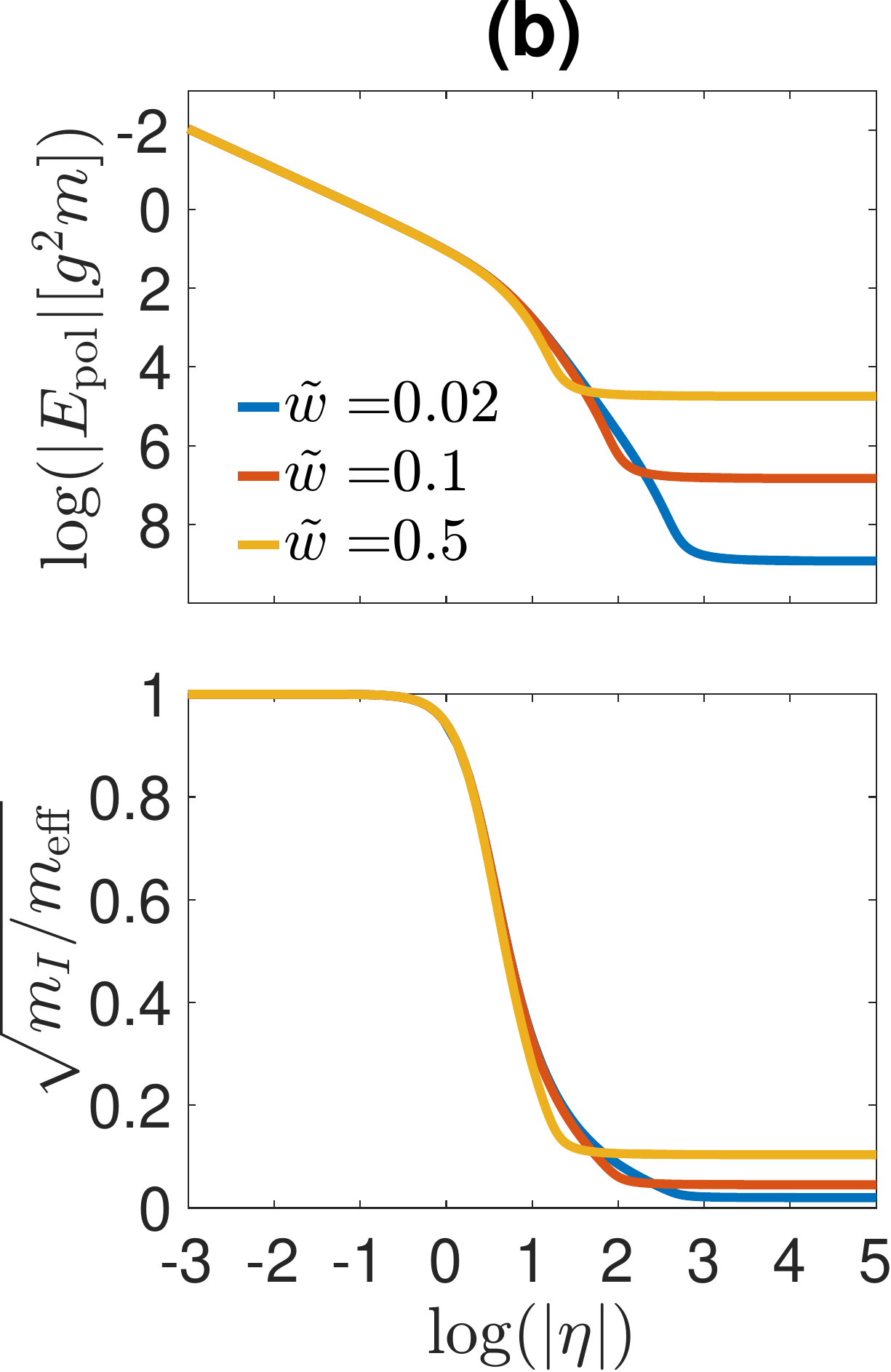}
    \end{subfigure}
    \caption{\justifying The polaron energy and the effective mass of the \textbf{(a)} repulsive, and \textbf{(b)} attractive square-well impurity for various values of the well size $w$ as a function of the coupling strength $\eta$.
    }
    \label{fig:various_w}
\end{figure}


Finally, we examine how the polaron properties depend on $w$ in Fig.~\ref{fig:various_w}. The polaron properties are largely insensitive to $w$ in the weak coupling regime $|\eta| \lesssim 3$. This persists for all $\eta$ for the repulsive impurity's energy. But deviations among the polaron properties begin to emerge at stronger couplings, leading to substantial differences in the saturated values of the effective mass, and even to orders-of-magnitude variation in the energy of the attractive polaron. 

\section{\label{sec: Universality}Polaron Properties in the Strongly Coupled Regime}

Having noted the strong dependence of polaron properties on interaction range in the strong-coupling regime, we now ask whether a description based on the length scales of the problem applies to a short --but finite-ranged-- impurity in this regime. To this end, we analyze $\phi_0(x)$ and $\phi_1(x)$ near the strongly coupled limit, taking the ratio of the range to the coherence length as a small expansion parameter, $\tilde{w} = w/\tilde{\xi} \ll 1$. This choice reflects the short-range nature of the impurity, enabling the derivation of analytical expressions in this regime. 

We first consider the solution of the repulsive impurity in the strong coupling limit $\aIB \to 0^-$. This occurs when $ \tilde{V_0} \simeq \left(\frac{z_r}{\tilde{w}/2}\right)^2$, where 
$z_r=\coth{z_r} \simeq 1.199$ (see Eq.~\eqref{aIB_w_V0_Rep}). 
In this limit,  $\phi_{0}(x)$  is approximately $\tanh{(|x|/(\tilde{\xi}\sqrt{2}))}$ for $|x|>{w}/2$ and $\cosh(\sqrt{W_r}x/\tilde{\xi})$ for $|x|<{w}/2$.  
Therefore, we can first determine the elliptic modulus $\nu_r = 1-\epsilon_r$, where $\epsilon_r \ll 1$, by solving the boundary condition equation \eqref{BCs_Repulsive2}: 
\begin{eqnarray}
    \epsilon_r \simeq \frac{1}{4 \mathrm{sinh}^2(u_r) W_r^2} = \frac{\tilde{w}^4}{64 \mathrm{sinh}^2(u_r)},
\end{eqnarray}
where $u_r \equiv \sqrt{W_r} \tilde{w}/2 \simeq z_r (1 + \mathcal{O}(\tilde{w}))$. Imposing continuity at $x = w/2$ yields: 
\begin{eqnarray}
    \frac{x_r}{\tilde{\xi}} = \frac{\coth^2(z_r)-1}{2z_r^2} \left( \frac{\tilde{w}}{2}\right)^3. 
\end{eqnarray}
Once $x_r$ is determined, the energy of the repulsive polaron can be approximated by evaluating the integral \eqref{pol_energy}, assuming $\phi_{0}(x) = \tanh((|x|+x_r)/(\sqrt{2}\tilde{\xi}))$ holds everywhere:
\begin{eqnarray}
    \frac{E_{\mathrm{pol}}}{gn_0^2 \tilde{\xi}} = \frac{4\sqrt{2}}{3} \left(1- \frac{3\sqrt{2}}{8}\frac{\coth^2(z_r)-1}{2z_r^2} \left( \frac{\tilde{w}}{2}\right)^3 \right). \nonumber \\
    \label{analytical_pol_Energy_Rep}
\end{eqnarray}
 The leading term corresponds to the energy of the dark soliton in the zero-range limit $\tilde{w} \to 0$, and the corrections appear at third order in $\tilde{w}$. This approximation is expected to slightly overestimate the polaron energy at larger values of  $\tilde{w}$, since the true solution inside the potential differs from the hyperbolic tangent.

\begin{figure}
    \begin{subfigure}{0.238\textwidth}
        \includegraphics[width = \textwidth,keepaspectratio]{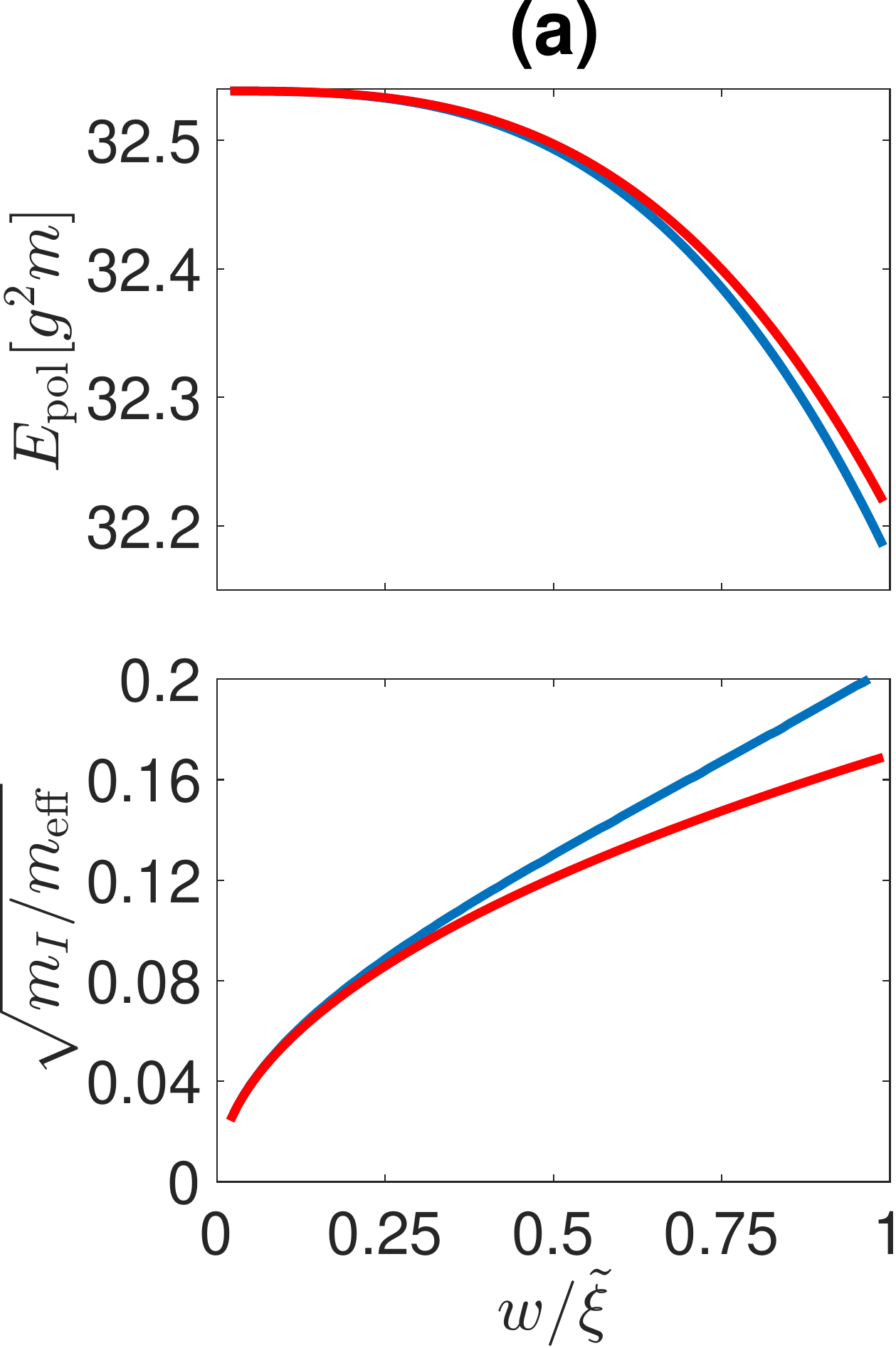}
    \end{subfigure}
    \begin{subfigure}{0.238\textwidth}
        \includegraphics[width = \textwidth,keepaspectratio]{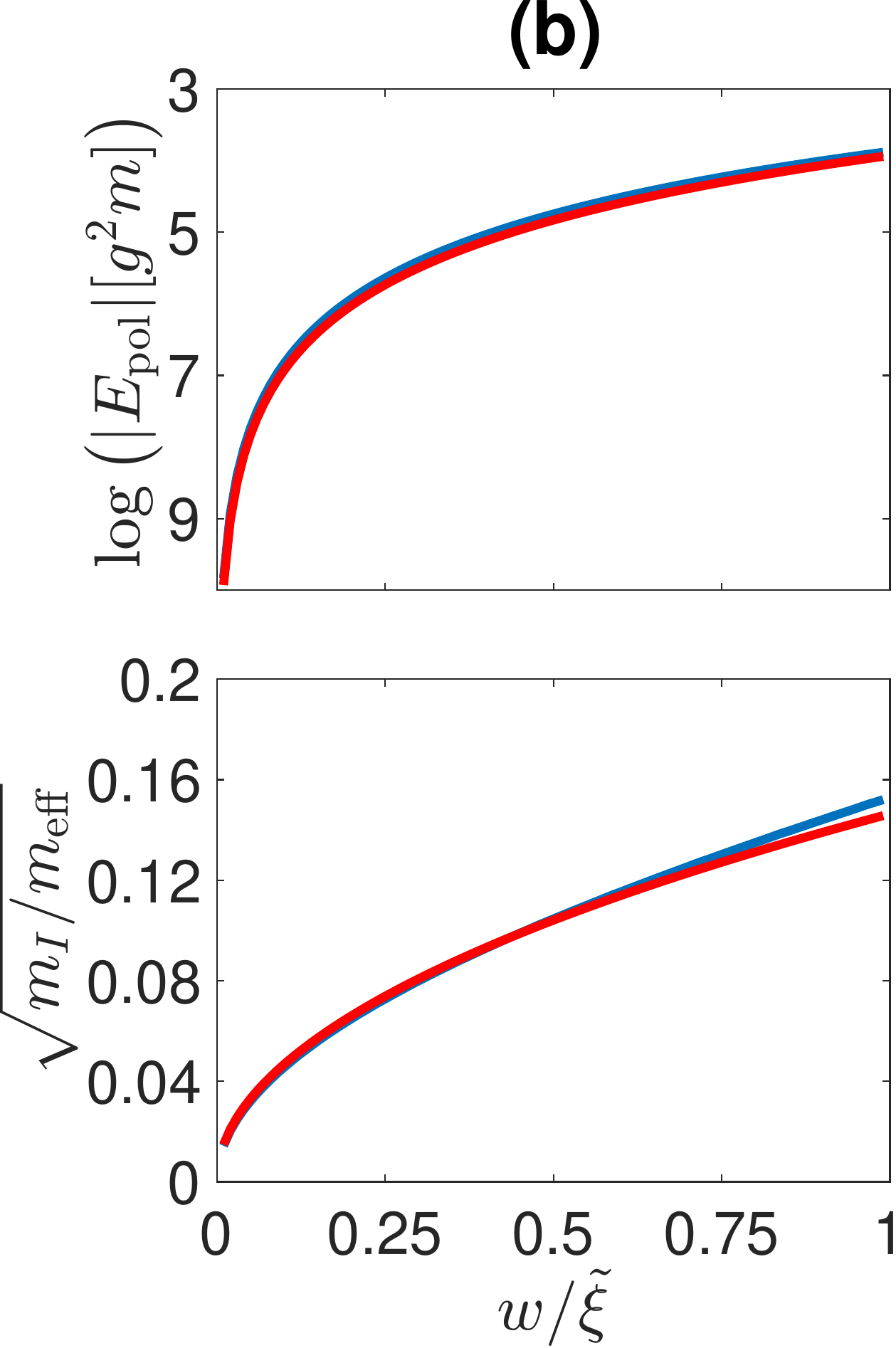}
    \end{subfigure}
    \caption{\justifying The polaron properties of \textbf{(a)} repulsive, and \textbf{(b)} attractive impurity at the strong coupling limit $\eta \to \infty$ as a function of the well size $w/\tilde{\xi} $. The blue lines represent the exact numerical results, while the red lines correspond to the analytical expressions derived from expansions valid in the limit of a small interaction range, $w/\tilde{\xi} \ll 1$. 
    }
    \label{fig:Universality}
\end{figure}
We derive an identity, valid for all $V_0$, specifically useful in determining the phase function $\phi_1(x)$ in the region $|x| \leq w/2$. One can obtain the expression $\partial_{xx}\phi_0/\phi_0 = \phi_0^2 -1 + \tilde{U}_{\mathrm{IB}}(x)$ from \eqref{phi0_equations}. Substituting this into \eqref{phi1_equations} and integrating by parts over the domain $x\in[0,\infty)$ yields:
\begin{eqnarray}
    \partial_x \phi_1(0) = \frac{\phi_0^2(0)-1}{2\phi_0(0)}.
    \label{eff_mass_origin_BC}
\end{eqnarray}
This result admits simple asymptotic approximations in the limits of strong impurity potentials. For a repulsive impurity, where $\phi_0(0)\ll 1$, $\partial_x \phi_1(0) \simeq -\frac{1}{2\phi_0(0)}$, while, for an attractive impurity, the expression simplifies to $\partial_x \phi_1(0) \simeq \frac{\phi_0(0)}{2}$.

Similar to the polaron energy, the dominant contribution to the effective mass of the repulsive polaron also originates from the region outside of the impurity. Noticing that $\partial_x \phi_{0} = \frac{1}{\sqrt{2}}(1-\phi_{0}^2)$ for $|x|>w/2$, we express the integral in the effective mass (Eq.~\eqref{pol_eff_mass}) as:
\begin{eqnarray}
    \int_{{w}/2}^{\infty}dx \phi_{1}(x)\partial_x\phi_{0}(x) = \frac{\partial_x \phi_1({w}/2)+1-\phi_0(w/2)}{\sqrt{2}}. \nonumber\\
    \label{eff_mass_identity}
\end{eqnarray}
 To compute the term $\partial_x \phi_1({w}/2)$, we note that $\phi_{0}(|x|<w/2) \simeq \sqrt{2W_r\epsilon_r} \cosh(\sqrt{W_r}x/\tilde{\xi}) + \mathcal{O}(\epsilon_r^{3/2})$. Therefore, the solution of Eq.~\eqref{phi1_equations} is
\begin{equation}
    \phi_1\left(|x|<w/2\right) \simeq -\frac{\sinh(z_r)}{\sqrt{2}}\sinh(\sqrt{W_r}x/\tilde{\xi}) + \mathcal{O}(\epsilon_r^{3/2}).
\end{equation}
Hence, the effective mass is 
\begin{eqnarray}
    \frac{m_{\mathrm{eff}}}{m_I} \simeq 2n_0 \tilde{\xi} \frac{m_r}{m_I}\frac{z_r \sinh (2z_r)}{\tilde{w}/2}.
    \label{analytical_rep_mEff}
\end{eqnarray}

For the attractive impurity in the strong coupling limit, we have $ \tilde{V_0} \simeq \left(\frac{z_a}{\tilde{w}/2}\right)^2$, where $z_a=-\cot z_a \simeq 2.798$ (see Eq.~\eqref{aIB_w_V0_Att}). The solution (Eq.~\eqref{wf_att_inside}) approaches the form
\begin{eqnarray}
    \phi_{0}(x) \simeq \sqrt{W_a}\left[1-\frac{\epsilon_a}{4} (\cosh(\sqrt{\frac{W_a}{2}}\frac{x}{\tilde{\xi}})-1)\right]
    \label{phi_in_approx}
\end{eqnarray}
for $|x|<w/2$, as the elliptic modulus satisfies $\nu_a = 1-\epsilon_a$ with $\epsilon_a \ll 1$. To second order in $\tilde{w}$, the boundary condition equation \eqref{BC_attractive} reads 
\begin{eqnarray}
    \epsilon_a \simeq \frac{2\left(1-\tanh(z_a/\sqrt{2})\right)}{1+\frac{1}{4}\left( \sqrt{2}z_a - \sinh (\sqrt{2}z_a)\right) \left( \tanh^2(z_a/\sqrt{2})-1\right)},\nonumber \\
\end{eqnarray}
which is only a constant at leading order, with corrections appearing at order  $\tilde{w}^2$. 
Once $\epsilon_a$ is known, the wave function is matched at $x=w/2$ in order to approximate $x_a$ to first order in $\tilde{w}$:
\begin{eqnarray}
    \frac{x_a}{\xi} = \frac{\tilde{w}}{2} \left(\frac{\sqrt{2}}{z_a} \frac{1}{1+ \frac{\epsilon_a}{4} \left(1- \cosh(\sqrt{2} z_a)\right) } -1\right).
\end{eqnarray}

The dominant contribution to the attractive polaron energy arises from the impurity region, since it is proportional to $\phi_0^4$. Using Eq.~\eqref{phi_in_approx}, the polaron energy $\propto \int_0^{w/2} (1-\phi_{in}^4)$ is
\begin{eqnarray}
    \frac{E_{\mathrm{pol}}}{gn_0^2 \tilde{\xi}} \simeq - \frac{z_a^4}{(\tilde{w}/2)^3} (1+ \mathcal{O}({\epsilon_a})),
    \label{analytical_pol_Energy_Att}
\end{eqnarray}
which, to leading order, scales as $\tilde{w}^{-3}$ due to the nearly flat density inside the impurity. 

The effective mass calculation requires integrating over all $x$, as both the interior and exterior regions contribute similarly. Hence, we evaluate the integral \eqref{pol_eff_mass} for both regions. We find $\phi_1(x) \simeq x \phi_0(x)/2$ for $|x|<w/2$ by solving \eqref{phi1_equations} with the boundary condition $\partial_x \phi_1(0) \simeq \phi_0(0)/2$, following from \eqref{eff_mass_origin_BC}, and substituting \eqref{phi_in_approx} for $\phi_0(x)$.  Then, the contributions from the exterior and interior regions can be calculated using \eqref{eff_mass_identity} and evaluating $\int_0^{w/2} x \phi_0 \partial_x \phi_0dx$, respectively. Combining these results yields
\begin{eqnarray}
    \frac{m_{\mathrm{eff}}}{m_I} \simeq 4n_0\tilde{\xi} \frac{m_r}{m_I} \left( \frac{\epsilon_a z_a^2 \cosh(\sqrt{2}z_a)}{4}+\frac{z_a}{\sqrt{2}}\right) \frac{2}{\tilde{w}},
    \label{analytical_att_mEff}
\end{eqnarray}
where the first (second) terms in the parentheses are contributions from inside (outside) the impurity range, and are approximately equal in magnitude.

In Fig.~\ref{fig:Universality}, we compare these expressions with the exact numerics. They show excellent agreement, with deviations appearing as expected when the ratio $w/\tilde{\xi}$ approaches unity. 
Both the polaron energy and the effective mass can be expressed solely as functions of the interaction range. For the attractive polaron, the energy scales as $\tilde{w}^{-3}$ at leading order, whereas in the repulsive case, the interaction range determines the subleading corrections—of order $\tilde{w}^3$—to the dark soliton energy. The effective mass for both types of impurities scales inversely with the interaction range, $m_{\mathrm{eff}} \propto \tilde{w}^{-1}$, with distinct prefactors, reflecting the inherent asymmetry between the two branches. 
\section{\label{sec: Conclusion} Conclusion}

We derived an analytical solution to the ground-state wave function of the GP equation for a one-dimensional Bose gas with a finite-ranged impurity and calculated perturbative corrections arising from the impurity’s slow motion. This approach resolves the divergence of polaron properties in the strongly coupled regime observed with contact potentials. Moreover, we found that in this regime the polaron properties admit a description determined by the ratio of the interaction range to the coherence length.
The developed formalism can also be applied to study polarons in the long-range regime, $\xi \gg w$, and to obtain excited-state solutions of the GP equation, enabling investigation of the connection between polaron branches and GP equation solutions in one dimension.

\acknowledgments
We gratefully acknowledge Nikolay Yegovtsev and Michael Fleischhauer for valuable discussions, and thank Gregory E. Astrakharchik and Fabian Grusdt for kindly providing the DMC data.
\bibliography{refs}

\appendix

\section{Scattering Length of the Finite-Well Impurity}\label{App:ScatteringLength}
\begin{figure}[t]
        \includegraphics[width=0.95\linewidth,keepaspectratio]{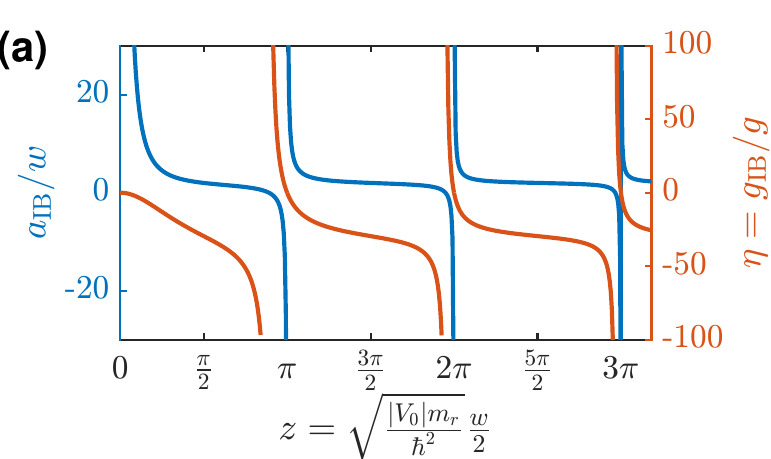}
        \includegraphics[width=0.95\linewidth,keepaspectratio]{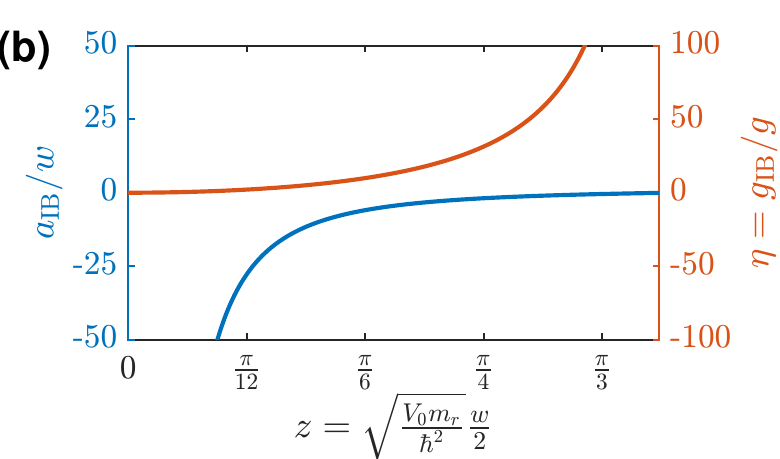}
    \caption{\justifying{The mapping between the parameters $z = \sqrt{\frac{|V_0|2m_r}{\hbar^2}} \frac{w}{2}$ of the square-well potential and the corresponding zero-energy scattering length $\aIB$ (blue lines), as well as the coupling constant $\gIB = -\frac{\hbar^2}{m_r\aIB}$ (red lines), is shown for \textbf{(a)} attractive, and \textbf{(b)} repulsive impurity. The gas parameter $\gamma = 1/(2n_0^2\xi^2)=0.438$, and the impurity-bath mass ratio $m_I/m = 0.47$ are used to obtain the coupling strength $\eta$.}}  
    \label{fig:aIB_gIB}
\end{figure}

Here, we briefly summarize the bound and scattering states of the square well potential. We restrict our analysis to the even-parity solution, since the ground-state wavefunction exhibits even parity. We illustrate the mapping between the square-well parameters, and the scattering length $\aIB$, Eq. \eqref{aIB_w_V0_Att} and Eq. \eqref{aIB_w_V0_Rep}, and the coupling strength $\gIB$ in Fig.~\ref{fig:aIB_gIB}.


An infinitesimally shallow attractive square well always supports an even-parity bound state. Additional odd- and even-parity states appear as $z$ crosses $(n-\tfrac12)\pi$ and $n\pi$, respectively, with $n$ a positive integer. The emergence of a new even-parity state coincides with a scattering-length divergence from $-\infty$ to $+\infty$. A repulsive square well supports no bound states and thus no such transition. Notably, the strongly coupled limit ($\gIB \to \pm \infty$) does not coincide with the unitarity points where new bound states form.

We model the repulsive impurity using a repulsive square-well potential ($V_0>0$) as a simplified representation of the impurity–bath interaction, interpreting the repulsive polaron as the ground-state solution under this potential. While this approach does not capture the metastable nature of repulsive polarons - typically decay into the ground state arising from attractive short-range interactions- it remains valuable for gaining physical insights into their behavior and exploring how their properties depend on the parameters of a finite-range potential. Analogous use of repulsive potentials to approximate short-range interactions between bath particles has yielded physically consistent results \cite{2020_PRR_Tilman_modified_GP_Bose_Polaron}.

\label{app:GPE}

\section{Properties of Jacobi elliptic functions} \label{Appendix:Jacobi}
The integral representations of Jacobi elliptic functions related to this work are as follows \cite{1948_Abramowitz_Mathematics,2013_Lawden_Elliptic_Functions}:
\begin{eqnarray}
    u &=& \int_1^{\mathrm{cd}(u,\nu)} \frac{dt}{\sqrt{(1-t^2)(1-\nu t^2)}}, \\
    u &=& \int_{\mathrm{dc}(u,\nu)}^1 \frac{dt}{\sqrt{(t^2-1)(t^2-\nu)}}, \\
    u &=& \int_1^{\mathrm{nc}(u,\nu)} \frac{dt}{\sqrt{(t^2-1)[(1-\nu)t^2+\nu]}},
\end{eqnarray}
where $\nu$ is the square of the elliptic modulus.
The standard Jacobi elliptic functions satisfy the identities:
\begin{eqnarray}
    \mathrm{sn}^2(u,\nu) + \mathrm{cn}^2(u,\nu) =1, \\
    \nu \mathrm{sn}^2(u,\nu) + \mathrm{dn}^2(u,\nu) =1.
\end{eqnarray}
Some useful derivatives and expansions are:
\begin{eqnarray}
    \frac{\partial \mathrm{sn}(u,\nu)}{ \partial  u} &=&  \mathrm{cn}(u,\nu) \mathrm{dn}(u,\nu),\\
    \frac{\partial \mathrm{cn}(u,\nu)}{ \partial  u} &=&  -\mathrm{sn}(u,\nu) \mathrm{dn}(u,\nu), \\
    \frac{\partial \mathrm{dn}(u,\nu)}{ \partial  u} &=&  -\nu \mathrm{sn}(u,\nu) \mathrm{cn}(u,\nu), 
\end{eqnarray}
\begin{eqnarray}
\mathrm{nc}(x,1-\epsilon)  &\simeq& \mathrm{cosh}(x)  \\
    &-& \frac{\epsilon}{8}[2x-\sinh(2x)]\sinh(x),  \nonumber \\
    \mathrm{cn}(x,1-\epsilon)  &\simeq& \mathrm{sech}(x)  \\
    &+& \frac{\epsilon}{8}[2x-\sinh(2x)]\tanh(x),  \nonumber \\
    \mathrm{sn}(x,1-\epsilon)  &\simeq& \mathrm{tanh}(x)  \\
    &+& \frac{\epsilon}{8}[2x-\sinh(2x)][\tanh^2(x)-1], \nonumber \\
    \mathrm{dn}(x,1-\epsilon)  &\simeq& \mathrm{sech}(x)  \\
    &+& \frac{\epsilon}{4} \mathrm{sech}(x) \bigg([\cosh(2x)-1] \nonumber \\ & & \ \ \ + \frac{1}{2}\tanh(x)[2x - \sinh(2x)]\bigg), \label{jacobi_expansions} \nonumber \\
    \mathrm{cd}(x,1-\epsilon)  &\simeq& 1+ \frac{\epsilon}{4} \left( 1- \cosh(2x)\right)     \label{cd_expansion}
\end{eqnarray}
\section{\label{Appendix:BC}The Equations of Boundary Conditions}

For the attractive well ($V_0<0$), the boundary conditions yield the transcendental equation:
\begin{eqnarray}\label{BC_attractive}
    0 &=&  \left( 1- \frac{2W_a \nu_a}{\nu_a+1}\right) - \frac{2W_a(\nu_a-1)\sqrt{\nu_a}}{\nu_a+1} \mathrm{sn}(u_a,\nu_a) \nonumber\\ 
    &+&\left( \frac{2W_a\nu_a}{\nu_a+1}-\nu_a\right)\mathrm{sn}^2(u_a,\nu_a),
\end{eqnarray}
where $u_a= \sqrt{\frac{W_a}{\nu_a+1}}\frac{w}{2\tilde{\xi}}$. The solutions are defined in the interval $\nu_a \in [0,1]$. 
For the repulsive well ($V_0>0$), the corresponding equations become:
\begin{eqnarray} \label{BCs_Repulsive1}
    0&=& \left( 1- \frac{2|W_r|}{1+\nu_r}\right) - \frac{2|W_r|(1-\nu_r)}{1+\nu_r} \mathrm{sn}(u_r,\nu_r) \nonumber \\ 
    &+& \left( \frac{2|W_r|\nu_r}{1+\nu_r}-1\right)\mathrm{sn}^2(u_r,\nu_r), \\
    0 &=&  \frac{2W_r(1-\nu_r)}{2\nu_r-1} + \frac{2W_r\sqrt{1-\nu_r}}{2\nu_r-1} \mathrm{sn}(u_r,\nu_r)\mathrm{dn}(u_r,\nu_r)\nonumber \\  &-&\mathrm{cn}^2(u_r,\nu_r), 
    \label{BCs_Repulsive2}
\end{eqnarray}
for $W_r<0$ and $W_r>0$, respectively. Here $u_r = \sqrt{\frac{|W_r|}{1+\nu_r}}\frac{w}{2\tilde{\xi}}$ ($u_r = \sqrt{\frac{W_r}{2\nu_r-1}}\frac{w}{\tilde{2\xi}}$) for $W_r<0$ ($W_r>0)$. The solutions are restricted to the intervals $\nu_r \in [0.5,1]$ for $W_r>0$, and $\nu_r \in [-1,1]$ for $W_r<0$.

\end{document}